\documentclass[prl,usletter,superscriptaddress,twocolumn]{revtex4}
\usepackage{amsmath}
\begin{document}
% You should use BibTeX and revtex.bst for references
%%\bibliographystyle{apsrev}

%\preprint{}

\title{Distinguishing separable and entangled states}


% \affiliation command. The \affiliation command MUST follow the

\author{A. C. Doherty}
%\thanks{}
\affiliation{Institute for Quantum Information, California Institute of
Technology }

\author{Pablo A. Parrilo}
\affiliation{Institut f\"ur Automatik, ETH Z\"urich}
\affiliation{Department of Control and Dynamical Systems, California
  Institute of Technology}

\author{Federico M. Spedalieri}
\affiliation{Institute for Quantum Information, California Institute of
Technology }


\date{\today}

\begin{abstract}
We show how to design families of operational criteria that
distinguish entangled from separable quantum states. The simplest of
these tests corresponds to the well-known Peres-Horodecki positive
partial transpose (PPT)
criterion, and the more complicated tests are strictly stronger. The
new criteria are tractable due to powerful computational and
theoretical methods for the class of convex optimization problems
known as semidefinite programs. We successfully applied the results to many 
low-dimensional states
from the literature where the PPT test fails. As a byproduct of the criteria, 
we provide an explicit
construction of the corresponding entanglement witnesses.
\end{abstract}

\maketitle

Entanglement is one of the most striking features of quantum mechanics. 
Not only is it at the heart of the violation of Bell inequalities~\cite{bell}, 
but it has lately  been recognized as a very useful
resource in the field of quantum information. Entanglement can be used
to perform several important tasks such as teleportation,
quantum key distribution and quantum computation~\cite{nielsen2000}.
Despite its widespread importance, there is not a procedure
that can tell us whether a given state is entangled or not, and
considerable effort has been dedicated to this problem in 
recent years~\cite{horodecki2001a,lewenstein2000b}. In this
letter we apply powerful tools of optimization theory for problems
known as semidefinite programs
to construct a hierarchy of tests that can detect entangled states.

A bipartite mixed state $\rho$ is said to be separable~\cite{werner1989a} 
(not entangled) if
it can be written as a convex combination of pure product states
\begin{equation}
\label{sep}
\rho =\sum p_{i}|\psi _{i}\rangle \langle \psi _{i}|\otimes |\phi
_{i}\rangle \langle \phi _{i}|,
\end{equation}
where $|\psi _{i}\rangle$ and $|\phi
_{i}\rangle $ are state-vectors on the spaces 
${\mathcal{H}}_{A}$\ and ${\mathcal{H}}_{B}$ of subsystems $A$ and $B$
respectively, and $p_i >0, \sum_i p_i =1$. If a state admits such a 
decomposition, then it can be created by local
operations and classical communication by the two parties, and hence it
cannot be an entangled state. 

Several operational criteria have been proposed to identify entangled
states. Typically these are based on simple properties obeyed by all
separable states and are thus necessary but not sufficient conditions
for separability (although some sufficient conditions for separability
are known~\cite{braunstein1999a}). The most
famous of these criteria is based on the partial transposition and
was first introduced by Peres~\cite{peres1996a}. It was shown by the
Horodeckis~\cite{horodecki1996a} to be
both necessary and sufficient for separability in ${\mathcal{H}}_{2}
\otimes {\mathcal{H}}_{2}$
and ${\mathcal{H}}_{2}\otimes {\mathcal{H}}_{3}$.
If $\rho$ has matrix elements 
$\rho_{i k,j l} = \langle i|\otimes \langle k| \rho |j \rangle \otimes
|l\rangle$
then the partial transpose $\rho^{T_A}$ is defined by
$\rho^{T_A}_{i k,j l}=\rho_{j k,i l}.$
If a state is separable, then it must have a positive partial
transpose (PPT). To see this consider the decomposition (\ref{sep}) for
$\rho$. Partial transposition takes $|\psi _{i}\rangle \langle \psi _{i}|$ 
to $|\psi_{i}^*\rangle \langle \psi_{i}^*|$, so the result of this
operation is another valid density matrix and must be positive.
Thus any state for which $\rho^{T_A}$ is not
positive semidefinite is necessarily entangled. This criterion has
the advantage of being very easy to check, but there are PPT states
that are nonetheless entangled as was first demonstrated
in~\cite{horodecki1997a}.

Our separability criteria will also be based on simple computationally
checkable properties of separable states. Consider the state
$\tilde\rho$ defined on  
${\mathcal{H}}_{A}\otimes {\mathcal{H}}_{B}\otimes {\mathcal{H}}_{A}$, given by
\begin{equation}
\label{ext}
\tilde{\rho}=\sum p_{i}|\psi _{i}\rangle \langle \psi _{i}|\otimes |\phi
_{i}\rangle \langle \phi _{i}|\otimes |\psi _{i}\rangle \langle \psi _{i}|.
\end{equation}
Firstly $\tilde{\rho}$ is an extension of $\rho $ (that is the partial trace
over the third party $C$ is equal to $\rho ,$ Tr$_{C}\left[ \tilde{\rho}
\right] =\rho $). Secondly the state is symmetric under interchanging the
two copies of ${\mathcal{H}}_{A}$. To put this more formally we define the
swap operator $P$ such that $P|i\rangle \otimes |k\rangle \otimes |j\rangle
=|j\rangle \otimes |k\rangle \otimes |i\rangle $. We have $P^{2}=I,$ and $
\pi =(I+P)/2$ is a projector onto the symmetric subspace. Since $\pi \tilde{
\rho}\pi =\tilde{\rho}$ the extension $\tilde{\rho}$ only has support on
this subspace. Finally the extension $\tilde{\rho}$ is a tripartite
separable state. This means that it will have positive partial transposes
with respect to any of the parties, and in particular we have
$\tilde{\rho}^{T_{A}} \geq 0$ and $\tilde{\rho}^{T_{B}}\geq 0$.

We may now formulate an explicit separability criterion based on the
existence of the extension discussed above. \emph{If the state }$\rho $ 
\emph{on }${\mathcal{H}}_{A}\otimes {\mathcal{H}}_{B}$\emph{\ is separable then
there is an extension }$\tilde{\rho}$\emph{\ on }${\mathcal{H}}_{A}\otimes 
{\mathcal{H}}_{B}\otimes {\mathcal{H}}_{A}$ \emph{such that }$\pi \tilde{\rho}
\pi =\tilde{\rho}$\emph{, }$\tilde{\rho}^{T_{A}}\geq 0$ \emph{and }$\tilde{
\rho}^{T_{B}}\geq 0$. Note that the symmetry of the extension means that if 
$\tilde{\rho}^{T_{A}}\geq 0$ then $\tilde{\rho}^{T_{C}}\geq 0,$ so including
this would not make a stronger test. We may generalize this
criterion to an arbitrary number of copies of both ${\mathcal{H}}_{A}$ and 
${\mathcal{H}}_{B}$. \emph{If the state }$\rho $ \emph{on }${\mathcal{H}}
_{A}\otimes {\mathcal{H}}_{B}$\emph{\ is separable then there is an extension }
$\tilde{\rho}$\emph{\ with support only on the symmetric subspace of }$
{\mathcal{H}}_{A}^{\otimes k}\otimes {\mathcal{H}}_{B}^{\otimes l}$ \emph{such
that }$\tilde{\rho}$\emph{\ has a positive partial transpose for all
partitions of the }$k+l$ \emph{parties into two groups.}\ Since the
extensions are required to be symmetric, it is only necessary to test the
possible partitions into two groups that are not related by permuting copies
of ${\mathcal{H}}_{A}$ and ${\mathcal{H}}_{B}$. Including testing for
positivity 
of the extension itself, there are $\lceil \left(
k+1\right) \left( l+1\right) /2\rceil $ distinct positivity checks to be
satisfied by $\tilde{\rho}$.  

These results generate a hierarchy of necessary conditions for 
separability. The first is the usual PPT test for a bipartite
density matrix $\rho$. If the test fails, the state is entangled; if
the test is passed, the state could be separable or entangled. In the
latter case we look for an extension $\tilde\rho$ of $\rho$ to three parties
such that $\pi \tilde\rho \pi = \tilde\rho$ that satisfies the PPT
test for all  
possible partial transposes.  If no such extension exists, then $\rho$ must 
be entangled. If such an extension
is possible, the state could be separable or entangled, and we
need to consider an extension to four parties and so on. 
 
Each test in this sequence is at least as powerful as the previous
one. We can see this by showing that if there is a PPT 
extension $\tilde\rho_{n}$ to $n$ parties, then there must be 
a PPT extension $\tilde\rho_{n-1}$ to $n-1$ parties. Let 
$\tilde\rho_{n-1} = \mathrm{Tr}_X [\tilde\rho_{n}]$, where $X$ represents one
of the copies of $A$ or $B$. It is easy to check that $\tilde\rho_{n-1}$
will inherit from $\tilde\rho_{n}$ the property of having its support on the 
symmetric subspace.  
Let's assume that it is not PPT. Then there
is a subset $\mathcal{I}$ of the parties such that 
 $\tilde\rho_{n-1}^{T_{\mathcal{I}}}$ has a negative eigenvalue, where $T_{\mathcal{I}}$ 
represents
the partial transpose with respect to all the parties in subset $\mathcal{I}$. Let
$|e\rangle$ be the corresponding eigenvector and
let $\{|i\rangle\}$ be a basis of the system $X$ over which the
partial trace was performed. Since $\tilde\rho_{n}$ is
PPT, then $\langle e|\langle i|\tilde\rho_{n}^{T_{\mathcal{I}}}|e\rangle|i\rangle \geq 0$,
for all $i$. Then
\begin{equation}
\label{rhonpt}
\sum_i  \langle e|\langle i|\tilde\rho_{n}^{T_{\mathcal{I}}}|e\rangle|i\rangle =
\langle e|{\mathrm{Tr}}_X [\tilde\rho_{n}^{T_{\mathcal{I}}}]|e\rangle \geq 0.
\end{equation}
Since $X\not\in \mathcal{I}$ , we can commute the trace and the partial
transpose, and using $\tilde\rho_{n-1} = \mathrm{Tr}_X
[\tilde\rho_{n}]$, we have 
$ \langle e|\tilde\rho_{n-1}^{T_{\mathcal{I}}}|e\rangle \geq 0$,
which contradicts the fact that $|e\rangle$ is an eigenvector of
$\tilde\rho_{n-1}^{T_{\mathcal{I}}}$ with negative eigenvalue. 

The problem of searching for the extension can be solved efficiently, since it 
can be stated as a particular case of the class of convex optimizations 
known 
as \textit{semidefinite programs} (SDP)~\cite{VaB:96}. A SDP 
corresponds to the
optimization of a linear function, subject to a linear matrix inequality
(LMI). A typical SDP will be 
\begin{eqnarray}
\label{sdp}
\text{minimize} &\ \  c^T {\bf x}  \nonumber \\
\text{subject to} &\ \  F({\bf x}) \geq 0, 
\end{eqnarray}
where $c$ is a given vector, ${\bf x} = (x_1,\ldots,x_m)$, and
$F({\bf x}) = F_0 + \sum_i x_i F_i$,
for some fixed $n$-by-$n$ hermitian matrices $F_j$. The inequality in
the second line
of (\ref{sdp}) means that the matrix $F({\bf x})$ is positive semidefinite. 
The vector ${\bf x}$
is the variable over which the minimization is performed. In the particular 
instance in which $c=0$, there is
no function to minimize and the problem reduces to whether or not it
is possible to find ${\bf x}$ such that $F({\bf x})$ is  
positive semidefinite. This is termed a feasibility problem. The
convexity of
SDPs has made it possible to develop
sophisticated and reliable analytical and numerical methods 
for them~\cite{VaB:96}.

The separability criteria we introduced above may all be formulated as
semidefinite programs. For brevity we will explicitly consider only the
problem of searching for an extension of 
$\rho$ to three parties. 
We will also relax the symmetry requirements on the
extension $\tilde\rho$, and we will ask only
$P\tilde\rho P=\tilde\rho$. This 
increases the size of the SDP, but simplifies the setup.
Let $\{ \sigma_i^A \}_{i=1,\ldots,d_A^2}, 
\{\sigma_j^B\}_{j=1,\ldots,d_B^2}$ be bases for the space of Hermitian
matrices that operate on ${\cal H}_A$ and  ${\cal H}_B$ respectively, 
such that they satisfy 
\begin{equation}
\label{tracesig}
\mathrm{Tr} ( \sigma_i^X \sigma_j^X) = \alpha \delta_{i j} \ \ \ \text{and}
\ \ \ \mathrm{Tr} (\sigma_i^X)  =  \delta_{i1},
\end{equation}
where $X$ stands for $A$ or $B$, and $\alpha$ is some constant---the
generators of $SU(n)$ could be used to form such a basis. 
Then we can expand $\rho$ in the basis 
$\{ \sigma_i^A \otimes \sigma_j^B \}$, and write
$\rho = \sum_{i j} \rho_{i j} \sigma_i^A \otimes \sigma_j^B$,
with $\rho_{i j}=\alpha^{-2} {\mathrm{Tr}} [\rho \, \sigma_i^A \otimes 
\sigma_j^B]$. We can write
the extension $\tilde\rho$ in a similar way
\begin{eqnarray}
\label{rhoext}
\tilde\rho & = & \sum_{\stackrel{i j}{\tiny{i<k}}}
\tilde\rho_{kji} \{\sigma_i^A 
\otimes \sigma_j^B \otimes \sigma_k^A +\sigma_k^A 
\otimes \sigma_j^B \otimes \sigma_i^A \} + \nonumber \\
& & + \sum_{kj}\tilde\rho_{kjk} \,\sigma_k^A 
\otimes \sigma_j^B \otimes \sigma_k^A, 
\end{eqnarray}
where we have explicitly used the symmetry between the first and third
party. We also need to satisfy $\mathrm{Tr}_C (\tilde\rho) = \rho$. 
Using (\ref{tracesig}), and the fact that
the $\sigma_i^A \otimes \sigma_j^B$ form a basis of the space of hermitian 
matrices on ${\cal H}_A \otimes {\cal H}_B$, we get $\tilde\rho_{ij1} 
= \rho_{ij}$.
The remaining components of 
$\tilde\rho$ will be the variables in our SDP. The LMIs come
from requiring that the state $\tilde\rho$ and its 
partial transposes be positive semidefinite. For example, the condition
$\tilde\rho \geq 0$ will take the form $F({\bf x}) = F_0 + \sum_i x_i F_i 
\geq 0$ 
if we define
\begin{eqnarray*}
\label{Fs}
F_0 & = & \sum_{j} \rho_{1j}\,\sigma_1^A \otimes\sigma_j^B \otimes\sigma_1^A 
+ \nonumber
\\
&&+ \sum_{i=2,j=1} \rho_{ij}\,\{\sigma_i^A \otimes\sigma_j^B \otimes\sigma_1^A +
\sigma_1^A \otimes\sigma_j^B \otimes\sigma_i^A\} \nonumber \\
F_{iji} & = &  \sigma_i^A \otimes\sigma_j^B \otimes\sigma_i^A 
\ \ \ \ \ \ \ \ \ \ \ \ \ \ \  i\geq 2,\nonumber \\
F_{ijk} & = &  (\sigma_i^A \otimes\sigma_j^B \otimes\sigma_k^A+
\sigma_k^A \otimes\sigma_j^B \otimes\sigma_i^A) 
 \ \ \ k> i\geq 2.
\end{eqnarray*}
The coefficients $\tilde\rho_{ijk} 
(k \neq 1, k\geq i)$ play the role of the variable
$\bf{x}$. There are $m=(d_A^4d_B^2-d_A^2d_B^2)/2$ components of $\bf{x}$, where
$d_I$ is the dimension of ${\cal H}_I$. Each $F$ is a square matrix of
dimension $n=d_A^2d_B$. 
Positivity of the partial transposes $T_A$ and $T_B$ leads to
two more LMIs, $\tilde\rho^{T_A} \geq 0$ and $\tilde\rho^{T_B} \geq
0$. The $F$ matrices for these two LMIs are related to the
matrices $F_{ijk}$ by the appropriate partial
transposition. 
We can write these three LMIs as one, if we define the matrix 
$G = \tilde\rho \oplus \tilde\rho^{T_A} \oplus \tilde\rho^{T_B}$, so for
example $G_0 = F_0 \oplus F_0^{T_A} \oplus F_0^{T_B}$ (a
block-diagonal matrix $C=A\oplus B$ 
is positive semidefinite iff both $A$ and $B$ are
positive semidefinite).  
So the feasibility problem reduces to attempting to find $\tilde\rho_{ijk}(k\neq 1, k \geq i)$
with $G \geq 0$.
In fact, the SDP corresponding to minimizing $t$ subject to $tI_{ABA}+G\geq 
0$ is always feasible and performs better
numerically. A positive optimum
gives a value of $p^*$
such that $(1-p)\rho+pI_{AB}/d_Ad_B $ is entangled for all $0\leq
p<p^*$. 
Looking for an extension on $
{\mathcal{H}}_{A}^{\otimes k}\otimes {\mathcal{H}}_{B}^{\otimes l}$
is a semidefinite program with $m=\left(\begin{smallmatrix}d_A+k-1 \\k
    \end{smallmatrix}\right)^2 \left(\begin{smallmatrix} d_B+l-1 \\l
    \end{smallmatrix} 
  \right)^2-d_A^2d_B^2 $ variables and a matrix  
$G$ with $\lceil \left(
k+1\right) \left( l+1\right) /2\rceil $ blocks of dimension at most
$\left(\begin{smallmatrix}d_A+\lceil k/2 \rceil-1 \\\lceil k/2 \rceil
    \end{smallmatrix}\right)^2 \left(\begin{smallmatrix} d_B+ \lceil l/2
\rceil -1 \\\lceil l/2
\rceil
    \end{smallmatrix} 
  \right)^2$.

Numerical SDP solvers are described in detail
in~\cite{VaB:96}. Typically they involve the solution of a series of
least squares problems each requiring a number
of operations scaling with problem size as $O(m^2n^2)$. For
SDPs with a block structure these
break into independent parts each with a value of
$n$ determined by the block size. The number
of iterations required is known to scale no worse than
$O(n^{1/2})$. Thus for any
fixed value of $(k,l)$ the computation involved in
checking
our criteria scales no worse than $O(d_A^{13k/2}d_B^{13l/2})$  which
is polynomial in the system size.  

Using
the SDP solver SeDuMi~\cite{SeDuMi}, we applied the first
criterion $(k=2,l=1)$ to several examples of 
PPT entangled states  with $d_A=2,d_B=4$ or $d_A=3,d_B=3$. On
a 500 MHz
desktop computer a single state could be tested in under a second for
$d_A=2,d_B=4$ and in around eight seconds for $d_A=3,d_B=3$. For the
one and two parameter families of PPT entangled states described
in~\cite{horodecki1997a,horodecki2001a,horodecki2000a} we performed a
systematic search of the parameter space, in each case testing
hundreds or thousands of different states. We checked 
4000 randomly 
chosen examples of the seven parameter family of PPT
entangled states states in~\cite{bruss2000a}. We also checked the
PPT entangled states constructed from unextendible product bases
in~\cite{bennett1999a}. We did not find any PPT entangled state
with an extension of the required form, thus verifying the entanglement
of all these states. Very close to the separable
states the test was
inconclusive due to numerical uncertainties. Uncertainties
and one example are discussed more fully below. 

A very useful property of a SDP, is the existence of the dual problem. 
If a problem can be stated as a SDP like (\ref{sdp}), usually called
the primal problem, then the dual 
problem corresponds to another SDP, that can be written
\begin{eqnarray}
\label{dual}
\text{maximize} &\ \ -\mathrm{Tr} [F_0 Z]  \nonumber \\
\text{subject to} &\ \  Z \geq 0 \nonumber \\
                       &\ \ \mathrm{Tr} [F_i Z] = c_i,
\end{eqnarray}
where the matrix $Z$ is hermitian and is the variable over which 
the maximization is
performed. For any feasible solutions of the primal and dual problems
we have
\begin{equation}
\label{certif}
 c^T {\mathbf{x}}+ \mathrm{Tr} [F_0 Z] = \mathrm{Tr} [F({\bf x}) Z]  \geq 0,
\end{equation}
where the last inequality follows from the fact that both $F({\bf x})$ and 
$Z$ are positive semidefinite. Then, for the particular case of a feasibility
problem ($c=0$), equation (\ref{certif}) will read $\mathrm{Tr} [F_0
Z] \geq 0$.
This result can be used to give a certificate of infeasibility for
the primal problem:  {\textit{if there exists
$Z$ such that $Z\geq 0$, ${\mathrm{Tr}} [F_i Z] = 0$, that satisfies ${\mathrm{Tr}} 
[F_0 Z] < 0$,
then the primal problem must be infeasible.}}

In the context of entanglement, the role of the ``certificate'' is
played by observables known as {\textit{entanglement witnesses}}
(EW)~\cite{horodecki1996a,terhal2000a}. 
An EW for a state $\rho$  satisfies
\begin{equation}
\label{ew}
\mathrm{Tr} [ \rho_{sep} W] \geq 0 \ \ \ \text{and} \ \ \ \text{Tr} [
\rho W] < 0,
\end{equation}
where $\rho_{sep}$ is any separable state. If our primal SDP is
infeasible (which means that the state $\rho$ must be entangled),
the dual problem provides a certificate of that infeasibility
that can be used to construct an EW for $\rho$.
 
First, we note that due to the block diagonal structure of the
LMI, we can restrict any feasible dual solution $Z$ to have 
the same structure,
i.e., $Z= Z_0 \oplus Z_1^{T_A} \oplus Z_2^{T_B}$
where the $Z_i$ are operators on ${\mathcal H}_A \otimes {\mathcal H}_B 
\otimes {\mathcal H}_A$. Then we have that
$\mathrm{Tr} [ G_0 Z] = \mathrm{Tr} [F_0 (Z_0+Z_1+Z_2)]$.
We defined $F_0$ as a linear function of $\rho$ so that
$F_0=\Lambda (\rho)$ where $\Lambda$ is a linear map from ${\mathcal
  H}_A \otimes {\mathcal H}_B$ to ${\mathcal H}_A \otimes {\mathcal
  H}_B \otimes {\mathcal H}_A $. We can now define an operator $\tilde{Z}$ on 
${\mathcal H}_A \otimes {\mathcal H}_B$ through the adjoint map
$\Lambda^*$ such that $\tilde{Z}=\Lambda^*(Z_0+Z_1+Z_2)$ and
\begin{equation}
\label{EW}
 \mathrm{Tr}[\rho \tilde{Z}] = \mathrm{Tr}[\Lambda(\rho) (Z_0+Z_1+Z_2)] =
 \mathrm{Tr}[G_0 Z].
\end{equation}

If $\rho_{sep}$ is any separable state, we know that the primal
problem is feasible (the extension $\tilde\rho$ exists). Then,
using $\mathrm{Tr} [G_0 Z] \geq 0$ and (\ref{EW}), we have $\mathrm{Tr} 
[\rho_{sep} \tilde{Z}] \geq 0$
for {\textit{any}} $\tilde{Z}$ obtained from a dual feasible
solution. For this particular problem, 
if the primal is not feasible (which means $\rho$ is an entangled
state), a feasible dual solution $Z_{EW}$ that satisfies
$\mathrm{Tr} [ G_0 Z_{EW}] < 0$ always exists.
Using (\ref{EW}) we can see that the corresponding operator 
$\tilde{Z}_{EW}$  satisfies
$\mathrm{Tr} [\rho \tilde{Z}_{EW}] < 0$
which together with $\mathrm{Tr} [\rho_{sep} \tilde{Z}_{EW}] \geq 0$ 
means 
that $\tilde{Z}_{EW}$ is an
entanglement witness for $\rho$.

In numerical work, if the SDP solver cannot find an extension
$\tilde{\rho}$ it constructs the matrices $Z_i$. Evaluating
$\mathrm{Tr}[\rho \tilde{Z}_{EW}]$ and verifying the three 
positivity conditions provides an {\it independent} check of the
result. Unless this check is not conclusive---for example,
if $\mathrm{Tr}[\rho \tilde{Z}_{EW}]$ is not significantly different
from zero---we 
are able to definitively conclude that no 
$\tilde{\rho}$ exists.  

If $W$ is an EW,  
then for any product state $|x y\rangle$ we have 
\begin{equation}
\label{bhform}
E(x,y)=\langle x y |W| x y \rangle = \sum_{ijkl} W_{ijkl} x^*_i y^*_j
x_k y_l \geq 0,
\end{equation}
where $\{x_i,y_i\}$ are the components of $|x\rangle, |y\rangle$ in some
basis, and $W_{ijkl}$ are the matrix elements of $W$ in the same
basis. Equation (\ref{bhform}) states that the 
biquadratic hermitian
form $E$ associated with $W$ must be positive semidefinite (PSD). 
It is not hard to show that all of the EWs
generated by Eqn.~(\ref{EW}) satisfy the relation
\begin{eqnarray}
\label{zext}
\langle x y x|\tilde{Z}_{EW}\otimes I| x y x\rangle&=& \langle x y x|(Z_0+Z_1+Z_2)| x y 
x\rangle \nonumber \\
&=&\langle x y x|Z_0| x y 
x\rangle +\langle x^* y x|Z_1^{T_A}| x^* y 
x\rangle \nonumber \\ && +\langle x y^* x|Z_2^{T_B}| x y^* 
x\rangle .
\end{eqnarray}
Since $Z_0,Z_1^{T_A}$ and $Z_2^{T_B}$ are positive by
construction the biquadratic hermitian form $E(x,y)
\langle x|x\rangle $ 
has a decomposition as a sum of squared magnitudes (SOS). This guarantees
that $E(x,y)$ is PSD. It can be
shown that our first separability criterion detects all entangled
states that possess an EW such that $E$ may be written in
this form. The dual program to our initial SDP may
be
interpreted as a search for an entanglement witness of this
kind.  
Equally, the Peres-Horodecki criterion detects the entanglement of
those states that possess entanglement witnesses for which (\ref{bhform})
may be written directly as a SOS---the decomposable entanglement
witnesses~\cite{lewenstein2000a} such that $W=P+Q^{T_A}$ for some PSD $P$
and $Q$. 
In general, if there is no
EW $W$ such that (\ref{bhform}) is a SOS, we can 
search over $W$ for which (\ref{bhform}) is a SOS when multiplied by
$\langle x|x\rangle^{k-1} \langle y|y\rangle^{l-1}$ 
for some $k,l \geq 1$. By duality, this corresponds to our $(k,l)$
separability 
criterion.

As an example illustrating the methodology, consider the state 
described in \cite[Section 4.6]{horodecki2001a}, given by:
\begin{equation}
\label{choiform}
\rho_\alpha = \frac{2}{7} | \psi_+ \rangle \langle \psi_+ | +
\frac{\alpha}{7} \sigma_+ +
\frac{5-\alpha}{7} P \sigma_+ P,   
\end{equation}
with $0 \leq \alpha \leq 5$, $|\psi_+\rangle = \frac{1}{\sqrt{3}} \sum_{i=0}^2 
|i i \rangle$,
$\sigma_+ = \frac{1}{3}(|0 1 \rangle \langle 0 1 | + 
|1 2 \rangle \langle 1 2 | + |2 0 \rangle \langle 2 0 |)$.
Notice that $\rho_\alpha$ is invariant under the simultaneous change
of $\alpha \rightarrow 5-\alpha$ and interchange of the parties. The
state is separable for $2
\leq \alpha \leq 3$ and not PPT for
$\alpha > 4$ and $\alpha < 1$. Numerically entanglement witnesses
could be constructed for
$\rho_{\alpha}$ in the range 
$3+\epsilon<\alpha\leq 4$ (and
$1\leq\alpha<2-\epsilon$) with $\epsilon\geq 10^{-8}$. A witness for
$\alpha>3$ can be extracted from these by inspection:
\begin{eqnarray*}
\tilde{Z}_{EW} &=& 
2 \, (|0 0 \rangle \langle 0 0 | +
   |1 1 \rangle \langle 1 1 | +
   |2 2 \rangle \langle 2 2 |)+ \\
&& + |0 2 \rangle \langle 0 2 | + 
|1 0 \rangle \langle 1 0 | + 
|2 1 \rangle \langle 2 1 | -
3 | \psi_+ \rangle \langle \psi_+ |.
\end{eqnarray*}
This observable is nonnegative
on separable states:
\[
\begin{array}{r}
2 \langle xy|\tilde{Z}_{EW}|xy \rangle \langle x|x \rangle  \quad = 
  |2 \, x_0 x_1 y_2^* - x_2 x_0 y_1^* - x_1 x_2 y_0^*|^2  \\
 +      | 2 x_0 x_0^* y_0 - 2 x_1 x_0^* y_1 + x_1 x_1^* y_0 - x_2 x_0^* y_2|^2  \\ 
 +      | 2 x_0 x_0^* y_2 - 2 x_1 x_2^* y_1 + x_2 x_2^* y_2 - x_0 x_2^* y_0|^2  \\
 +      | 2 x_0 x_1^* y_0 - 2 x_2 x_2^* y_1 + x_2 x_1^* y_2 - x_1 x_1^* y_1|^2  \\
 + 3 \, | x_2 x_0 y_1^* - x_1 x_2 y_0^*|^2  
 + 3 \, | x_1 x_1^* y_0 - x_2 x_0^* y_2|^2  \\ 
 + 3 \, | x_2 x_2^* y_2 - x_0 x_2^* y_0|^2  
 + 3 \, | x_2 x_1^* y_2 - x_1 x_1^* y_1|^2 \geq  0. 
\end{array}
\]
The expected value on the original state is $
\mbox{Tr}[\tilde{Z}_{EW} \rho_\alpha] = \frac{1}{7} (3-\alpha)$, demonstrating
entanglement for all $\alpha > 3$.

The reformulation of our separability tests as a search for SOS
decompositions of the forms $E(x,y)$ provides connections with
existing results in real algebra (see \cite{parrilo2001a} for a
discussion of the SDP-based approach in a general setting). By Artin's
positive solution to Hilbert's 17th problem, for any real PSD form
$f({\mathbf{x}})$ there exists a SOS form $h({\mathbf{x}})$, such that the
product $f({\mathbf{x}}) h({\mathbf{x}})$ is
SOS~\cite{reznick2000a}. Finding such an $h({\mathbf{x}})$ and SOS
decomposition proves that $f$ is PSD.  For a fixed SOS form
$h(x,y)$, we may write a SDP that attempts to find EWs such that
$h(x,y) E(x,y)$ is SOS. In our hierarchy of criteria the form $h$ is
restricted to be $\langle x|x\rangle^{k-1} \langle y|y\rangle^{l-1}$. While it
is conceivable that every PSD bihermitian form is SOS when multiplied
by appropriate factors of this kind, currently we do not have a proof. 
It is known that deciding whether a form is positive is NP-hard and so
this connection to positive forms also promises to shed light on the
computational complexity of the separability problem.

In this letter we introduced a hierarchy of separability tests
that are computationally tractable and strictly stronger than the
PPT criterion. Only the second step in this sequence
of tests was required to detect the entanglement of a wide
class of known PPT entangled states. The method is based on
the application of semidefinite
programs. By exploiting the duality property of these problems, 
we showed how to construct entanglement witnesses for states
that fail any separability test in the sequence. 
These numerical results can also be very helpful in finding 
analytical expressions for the entanglement witness.
The application of this approach to the
characterization of positive maps will be
reported elsewhere. Finally, the wide
range of applications of semidefinite programming, along with the work
reported here and in~\cite{rains2001a},
suggests that it may become a useful tool in 
quantum information and in quantum theory in general. 

It is a pleasure to acknowledge conversations
with Hideo Mabuchi, John Doyle, John Preskill and Patrick Hayden. This work
was supported by the Caltech MURI Center for Quantum
Networks, the NSF Institute for Quantum Information and the Caltech
MURI Center for Uncertainty Management for Complex Systems.

\end{document}